\documentclass[twocolumn,showpacs,preprintnumbers,amsmath,amssymb]{revtex4}
\usepackage{graphicx}
\usepackage{dcolumn}
\usepackage{bm}
\begin{document}
\preprint{UNLV-HiPSEC-M03-01}
\title{Competition between phase coherence and correlation in a mixture
of Bose-Einstein condensates}
\author{Hong Ma and Tao Pang}
\affiliation{Department of Physics, University of Nevada, Las Vegas,
Nevada 89154-4002}
\date{\today}
\begin{abstract}
Two-species hard-core bosons trapped in a three-dimensional isotropic harmonic
potential are studied with the path-integral quantum Monte Carlo simulation.
The double condensates show two distinct structures depending on how the
external potentials are set.  Contrary to the mean-field results, we find that
the heavier particles form an outer shell under an identical external potential
whereas the lighter particles form an outer shell under the equal energy
spacing condition.  Phase separations in both the spatial and energy spaces are
observed.  We provide physical interpretations of these phase separations and
suggest future experiment to confirm these findings.
\end{abstract}
\pacs{03.75.Mn, 05.30.Jp, 02.70.Ss}
\maketitle
The rich physics of multi-species boson mixtures has attracted much attention
since the first realization of the coexistence of the double condensates from
two different spin states $|F=2,m=2\rangle$ and $|F=1, m=-1\rangle$ of
$^{87}$Rb~\cite{my97}.  Some interesting properties, such as the stability of
condensates, phase separation, and symmetry breaking, have been investigated
both experimentally~\cite{my97,mo02} and
theoretically~\cite{ho96,pu98,ba97,es97,oh98,ao98,ti98,go98,ka01,ri02}.
It was found that $^{87}$Rb in two hyperfine states can form two separate
condensates with a spatial overlap~\cite{my97} and this overlapping region can
be adjusted by tuning the interspecies interaction.  Theoretical
studies including both the Thomas--Fermi approximation and numerical
solution of the Gross--Pitaevskii equation~\cite{ho96,pu98,oh98,ri02} have also
found that two overlapping condensates
can be separated into a core at the trap center and a surrounding shell.
This phase separation can be achieved by changing the interspecies interaction,
the total number of particles, and the symmetry of the trapping potential.
It is suggested that the system may become unstable for certain
interaction strength~\cite{pu98}, which may lead to a spontaneous
symmetry breaking spatially~\cite{es97,oh98,ao98}.

Most of the early theoretical efforts mentioned above only give the
properties of condensates at zero temperature.  In this Letter, we study
a two-species boson system in a three-dimensional isotropic harmonic trap 
through the path-integral quantum Monte Carlo simulation, which allows us to
probe such a system at finite temperature.  Here we
concentrate on the conditions under which the phase separation occurs. 

The path-integral quantum Monte Carlo method has been successfully applied
to the one-species boson systems in traps~\cite{pe98}.  It provides the exact
solution of a many-body system within a controllable variance.  The total
energy, density profile, specific heat, condensation fraction, and other
quantities can all be evaluated~\cite{pe98}. 

The system considered here consists of two types of particles, marked
as species 1 and species 2, respectively, with total particle
numbers $N_{1}$ and $N_{2}$ and particle masses $m_{1}$ and $m_{2}$.
For a realistic system, these two types of particles can be two different
elements, two different isotopes of the same element, or two different
hyperfine states of identical atoms.  The interactions between two like
and unlike atoms are characterized by the $s$-wave scattering lengths
$a_{11}$, $a_{22}$, and $a_{12}$.  In the model studied here, $a_{11}$ and
$a_{22}$ are the hard-core diameters of the particles in species 1 and 2,
respectively, and $a_{12}=(a_{11}
+a_{22})/2$.

The Hamiltonian for the system of $N=N_1+N_2$ particles is given by
\begin{equation}
\mathcal{H}=\mathcal{H}_1+\mathcal{H}_2+\sum_{j>k=1}^N V_{jk},
\end{equation}
where
\begin{equation}
\mathcal{H}_{i}=-\frac{\hbar^2}{2m_i}\sum_{l=1}^{N_{i}}\nabla_l^2
+\sum_{l=1}^{N_{i}} U_i(r_l)
\end{equation}
is the Hamiltonian of the noninteracting system of the $i$th species under
the spherically symmetric trapping potential $U_i(r)=k_ir^2/2=m_i\omega_i^2
r^2/2$.  The interaction between any two particles $V_{jk}$ is assumed to be a
hard-core potential.

The algorithm applied to the one-species systems~\cite{pe98} is generalized
to study the two-species system here.  We treat the two-species system as
two subsystems, each of which contains one species and follows its own
statistics. The permutations are only performed among the identical particles
during a simulation.  However, these two subsystems do not behave independently
because of the interspecies interaction. The Monte Carlo steps are influenced 
by the interactions among all the particles as well as the permutations of 
the particles in each species.

In order to calculate the density profile, we divide the space into small
shells along the radial direction.  We choose the interval between two shells
such that after averaging over a large number of Monte Carlo steps on the
order of $10^{6}$, the finite-size error diminishes in comparison with
the statistical error.  The spatial density $\rho_i(r)$ can then be obtained
by counting the number of particles from each species in a given shell.
The normalization relation $\int\rho_i(r)\,d\mathbf{r}=N_i$ is used to check
the convergence.
For convenience, we use the dimensionless units with $\hbar=m_1=\omega_1=1$.

We start the simulation with two species with an identical mass, hard-core
radius, and total particle number to verify that the condensation does occur
below a certain temperature.  The two condensates
show different critical temperatures as we change the ratios $a_{11}/a_{22}$
and $m_2/m_1$. The stronger the interaction is, the lower the critical
temperature.  The condensation favors the lighter particles because a heavier
particle system has a lower critical temperature, which agrees well with the
available experimental results~\cite{mo02}. 

Our main interest here is on the structures of the mixtures.  We calculate the
densities of two condensates at various parameters at temperature
$T/T^{0}_{\mathrm{c}}=0.1$ when both the species have undergone Bose--Einstein
condensation.  Here $T^0_{\mathrm{c}}$ is the condensation/critical temperature
under the mean-field approximation~\cite{pe98} for species 1 with a total
number of particles $N$.  The mixtures
possess two distinct structures depending on the relationship between the
external potentials on the two species.  We have examined two cases, one with
an equal energy spacing in the single-particle spectra and another with an  
identical external potential.  We will elucidate the results for both cases.
 
\begin{figure}[!hb]
\includegraphics[width=8.2cm,height=9.7cm]{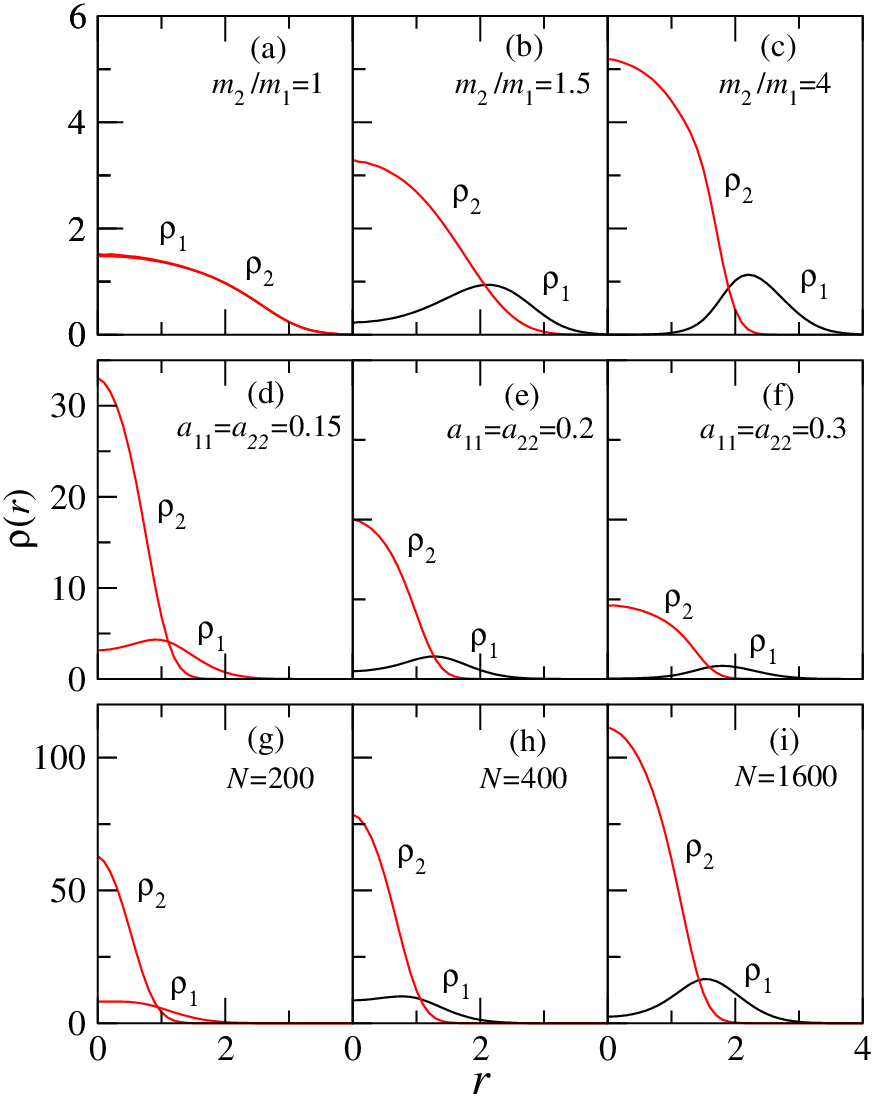}
\caption{\label{fig1}Density profiles of the double condensates with various
parameters at $T/T^{0}_{\mathrm{c}}=0.1$ and with $N_1=N_2$. (a)--(c) show
the changes of the density profiles at different mass ratios but with
$N=200$ and $a_{11}=a_{22}=0.4$. (d)--(e) show the cases at different
interaction strengths with $m_2/m_1=4$ and $N=200$. (g)--(i) show the
dependence of the total particle number with $m_2/m_1=4$ and $a_{11}
=a_{22}=0.1$.}
\end{figure}

We first consider the case with the same energy spacing
in the single-particle spectra, that is, $\omega_1=\omega_2$.  Two identical
condensates are found when $m_2/m_1=1$, $N_2/N_1=1$, and $a_{11}/a_{22}=1$,
as shown in Fig.~\ref{fig1}(a).  Comparing with the mean-field result on the
one-species condensate~\cite{da99}, the double condensates in our simulation
display the same expansion as $Na_{jk}$ is increased from 2 to 40.

In order to separate the two overlapped condensates, we perform simulations
with various parameters $m_2/m_1$, $a_{11}=a_{22}$, and $N_2=N_1$, respectively.
Figures~\ref{fig1}(a)--(i) show the density profiles of the two species in the
system of these cases.  There are two significant features in the density
profiles as $m_2/m_1$ is increased from 1 to 4, as shown in
Figs.~\ref{fig1}(a)--(c): The heavier species remains to be concentrated
at the central area of the trap while the lighter particles are pushed outward
and finally form an outer shell surrounding the heavier species.

The interaction dependence of the density profiles is given in
Figs.~\ref{fig1}(d)--(f). The interaction strength can dramatically affect
the particle distribution.  Here we show the results for
the same number of particles in each species with $m_2/m_1=4$.  The densities
of both the species at the trap center decrease as the interaction strength
is increased from $a_{ij}=0.15$ to 0.3.  However, under a stronger interaction
strength, the lighter particles are pushed out further away from the trap
center.  Heavier particles spread out with a less distance.

Figures~\ref{fig1}(g)--(i) show the changes in the density distribution as the
total number of particles increases from 200 to 1600 while keeping $N_2=N_1$,
$m_2/m_1=4$, and $a_{11}=a_{22}=0.15$.  The two condensates are always in
partial overlap but with more the lighter particles going outward with a larger
$N$.  This means that when more particles are trapped in an
experiment, the heavier particles can be caged inside the light particles
with both the species forming condensates at the low energy levels of the
single-particle spectra.

The condensation fraction in a species can be determined by counting the
particles on the permutation rings~\cite{pe98}.  We have evaluated the
condensation fractions for each species in the system with parameters
shown in Fig.~\ref{fig1} and find them in the range of 0.8--0.95.
This means that a reasonable fraction of particles is not in phase coherence
with the rest in that species.  Furthermore, particles can accumulate in any
single-particle state, for example, the first excited state, to form a
condensate.  In fact, in certain parameter regions, the particles can be
completely driven out of the single-particle ground state by strong correlation.

\begin{figure}[!ht]
\includegraphics[width=8.4cm,height=11.8cm]{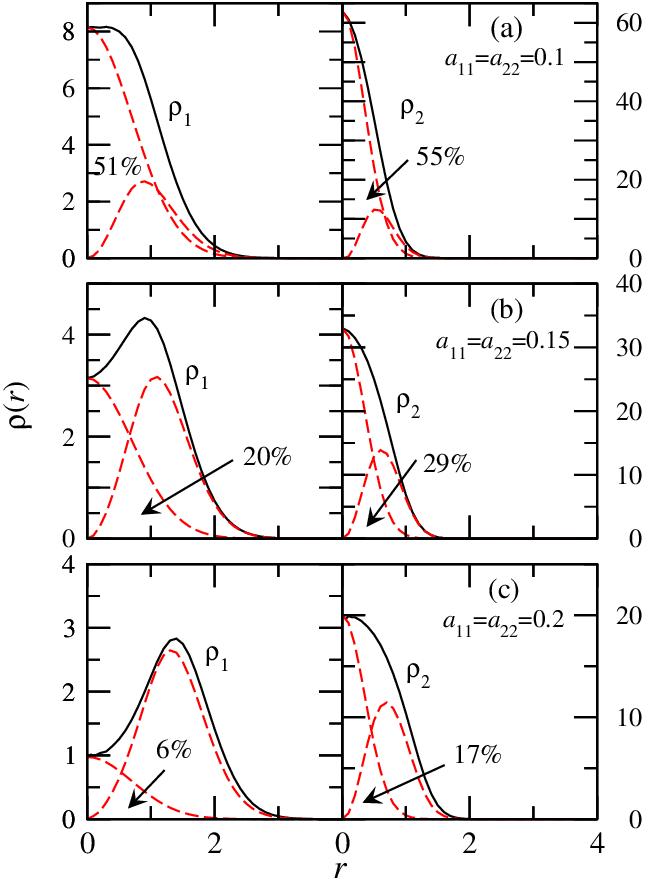}
\caption{\label{fig2}Decomposition of the density of each species into
that of the single-particle ground state and the corresponding excited
states at different interaction strengths with $m_2/m_1=4$ and $N_1=N_2=100$.
The percentage in each plot indicates the fraction of the particles in the
single-particle ground state.}
\end{figure}

To have a better picture of whether a particle is in its single-particle ground
state or not, we further decompose the density of each species into that of
the single-particle ground state and the corresponding excite states in
Fig.~\ref{fig2}.

We show here the percentages of the particles in each species in the
single-particle ground states at different interaction strengths with
$m_2/m_1=4$ and $N_1=N_2=100$.  The percentage of the particles in the
single-particle ground state drops as the interaction strength is increased
for both the species.  This can be interpreted as a result of competition
between correlation and phase coherence.  When the strength is small
(with small hard-core radii), the particles are close to be free and they
condense to the single-particle ground state when the temperature is lowered
beyond the critical temperature.  However, when the interaction strength is
increased, particles try to avoid each other to lower the total energy of
the system, and more particles move into
the low-lying excited states with higher single-particle energies.  
Our simulation shows that the lighter species is affected more drastically 
by the interaction
strength than the heavier one because the corresponding percentage drops
faster as the interaction strength is increased.

Decomposition of the density profiles indicates that when the spatial
separation between the two species occurs, the majority of particles
from both the species do not occupy the single-particle ground states.
This is because when one species is pushed away from the central area
of the trap by the correlation, the particles in both the species are
also driven away from the single-particle ground states by the same force. 
The distribution of the particles in the energy space is strongly influenced
by the energy spacing and the scattering lengths.  Therefore, no significant
separation between the two species can happen in the energy space when
$\omega_1=\omega_2$ and $a_{11}=a_{22}$.

\begin{figure}[!ht]
\includegraphics[width=8.2cm,height=5.2cm]{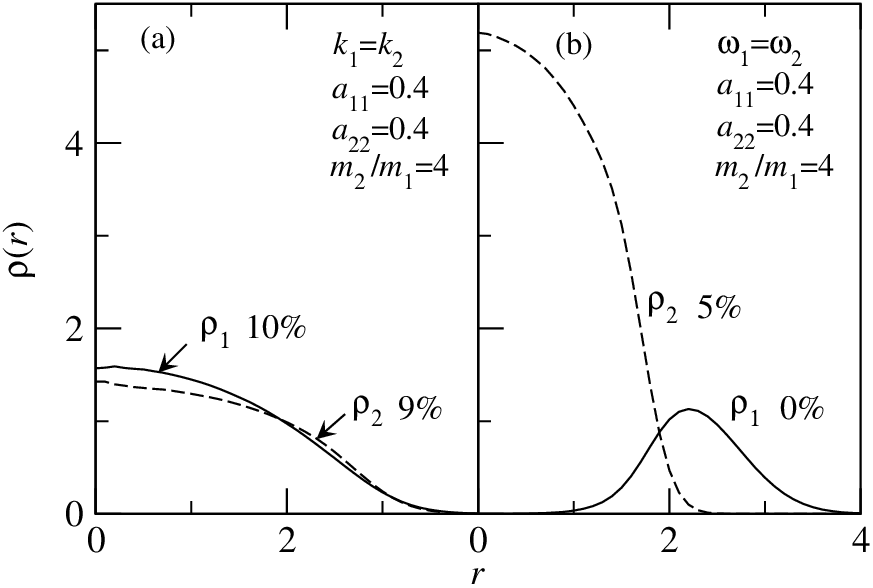}
\caption{\label{fig3}Density profiles of two-species system with the identical
external potential and equal energy spacing.  The percentage in each plot
indicates the fraction of particles in the single-particle ground state.  A
shell of lighter particles is formed under the equal energy spacing condition.}
\end{figure}

So far we have shown that the lighter particles can form an outer shell
surrounding the heavier particles.  A mass ratio far away from 1, strong
interaction, and large number of particles are all the contributing factors
for the particles to accumulate in the higher single-particle energy states
and to cause a spatial separation between the two species.  However, the
structure of a mixture turns out to be very different if the external
trapping potentials on the two species are different from those with
$\omega_1=\omega_2$.  Let us consider the case with an identical external
potential, that is, with $k_1=m_1\omega_1^2 =k_2=m_2\omega_2^2$. In 
Fig.~\ref{fig3}, we compare the results under different external potentials.  
The percentage indicates the fraction of particles in the single-particle 
ground state.  The simulation result shown in Fig.~\ref{fig3}(a) under an 
identical external potential does not have any significant phase
separation even though most particles are pushed out of the single-particle
ground states.  While keeping the same energy spacing, the spatial phase
separation between the two species is obvious, as shown in Fig.~\ref{fig3}(b).

\begin{figure}[!ht]
\includegraphics[width=8.2cm,height=5.2cm]{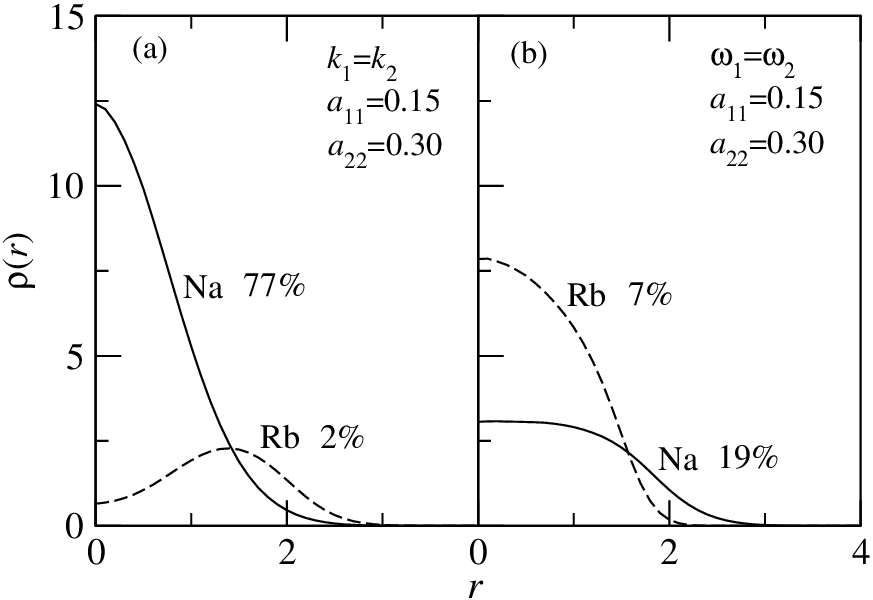}
\caption{\label{fig4}Density profiles of Rb and Na with the identical
external potential and equal energy spacing.  The percentage in each plot
indicates the fraction of particles in the single-particle ground state.
Separations in both the spatial and energy spaces occur with the identical
external potential.}
\end{figure}

More interestingly, the roles of the heavier and lighter species can be reversed
under the identical potential case with the heavier particles forming an
outer shell surrounding the lighter particles. To consider a possible 
experiment, we have performed the simulation with parameters appropriate to the
$^{87}$Rb--$^{23}$Na mixture.  We set Na as species 1 and Rb as species 2.
The total particle numbers $N_1=N_2=100$ and the hard-core radii used in
producing the results shown in Fig.~\ref{fig4}(a) are equivalent to an actual
mixture with $s$-wave scattering lengths of 3~nm and 6~nm, respectively, and a
total particle number on the order of $10^{4}$~\cite{pu98}.
In Fig.~\ref{fig4}, the density profiles of both the species under the same
single-particle energy spacing and identical external potential are shown.
The energy spacing in the single-particle spectrum of Rb is about half of
that of Na for the data shown in Fig.~\ref{fig4}(a) due to the difference
in the trapping frequencies.  The small energy gap and large scattering
length make it much easier for Rb atoms to be driven out of the trap center
or its single-particle ground state whereas most Na atoms stay at the trap
center or in its single-particle ground state. This is totally
counterintuitive and in disagreement with the mean-field
results~\cite{ho96,pu98,es97}. No significant spatial
separation is found when the single-particle energy spacings are the same,
as shown in Fig.~\ref{fig4}(b); but increasing either the interaction
strength or the total number of particles can result in the spatial
separation with an outer shell of Na.

Under the identical external potential, we
have also found that the spatial separation of the two species is independent
of the ratio $m_2/m_1$.  Therefore, for a system containing any two species,
it is energetically favorable to have the species with a larger scattering
length to form a low-density outer shell surrounding the other with a smaller
scattering length; this is consistent with the experimental observation for the
mixture of $^{87}$Rb atoms with two distinct spin states~\cite{my97}.  It is
also possible for a future experiment to realize the separation in the energy
space as predicted here.

Our simulations show that the external potentials can be used to tune the
structure of a boson mixture.  With the same trapping frequency for the two
species, it is more likely for a system with a mass ratio close to 1 to form
an outer shell of particles with a larger scattering length whereas for system
with a mass ratio far away from 1 to form an outer shell of lighter particles.
In the latter case, the large difference of external potentials between two
species outweighs the moderate effect of the two-body interactions within a
species.  It is possible for the $^{87}$Rb--$^{41}$K mixture to form an outer
shell of Rb atoms whereas for the $^{87}$Rb--$^{23}$Na mixture to form an outer
shell of Na atoms and the future experiment will tell.
 
We thank Dr.\ Han Pu for providing helpful information and comments. This work
was supported in part by the U.S. Department of Energy under the Cooperative
Agreement DE-FC08-01NV14049-A09.
\end{document}